\documentclass[10pt]{article}

\usepackage[pdftex]{graphicx}

\DeclareGraphicsExtensions{.pdf,.mps,.png,.jpg,.gif}

\begin{document}

\noindent
{\bf \large QCD and Hadronic Interactions with Initial-State-Radiation at B-Factories}

~

\noindent
D.~Bernard
\footnote{From the BaBar Collaboration
\hfill 
email: denis.bernard at in2p3.fr}

\noindent
{\em
Laboratoire Leprince-Ringuet - Ecole polytechnique, CNRS/IN2P3, 91128 Palaiseau, France
}

~

{\bf Summary.} --- 
The efforts to improve on the precision of the measurement and 
theoretical prediction of the anomalous magnetic moment of the muon
 $a_\mu$
have turned into a test of our understanding of the hadronic
contribution to vacuum polarisation.

I describe how recent measurements of hadron production in
 $e^{+}e^{-}$ interactions with initial-state radiation provide
 precision measurements of the hadron cross section, and have improved on the contribution to the prediction of the value of 
 $a_\mu$ that dominates the global uncertainty.

~

\noindent
PACS 13.66.Bc -- Precision measurements in $e^{+}e^{-}$ interactions \\
PACS 12.38.Qk -- Quantum chromodynamics (Experimental tests) \\
PACS 13.40.Em -- Electric and magnetic moments

\section{Introduction} 

\begin{figure}[t]
 \begin{center}
 \includegraphics[width=0.32\textwidth]{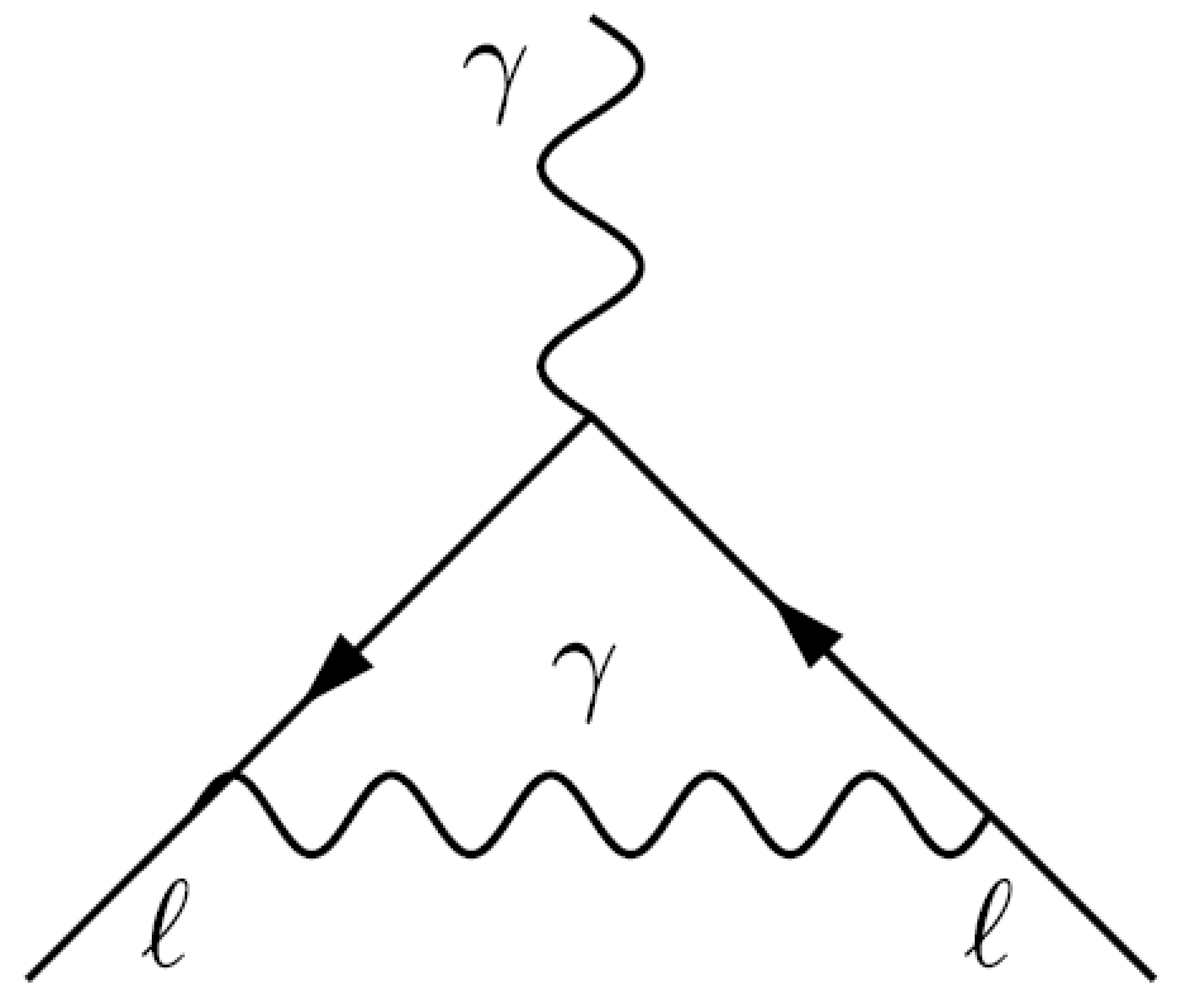}
 \includegraphics[width=0.32\textwidth]{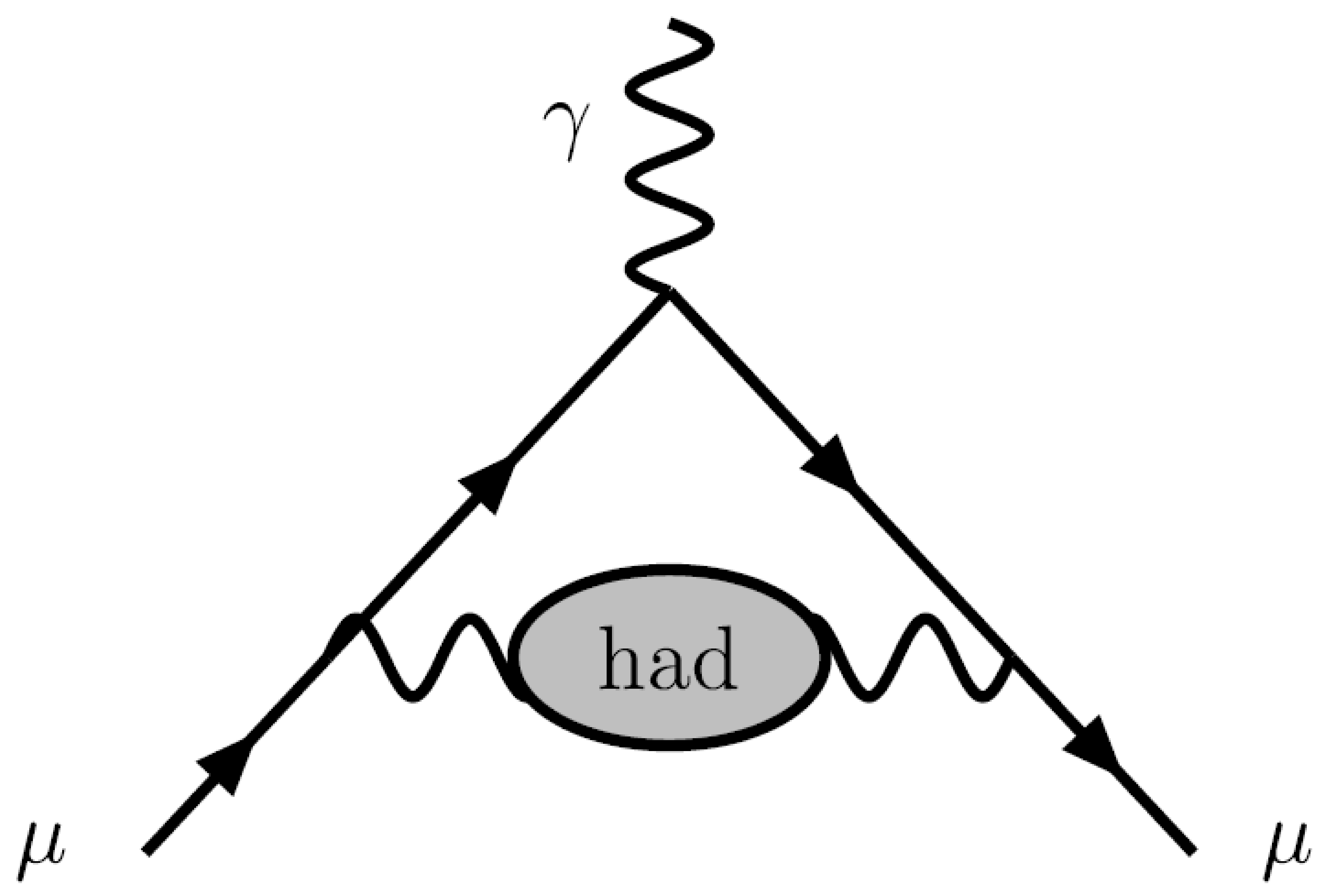}
 \includegraphics[width=0.32\textwidth]{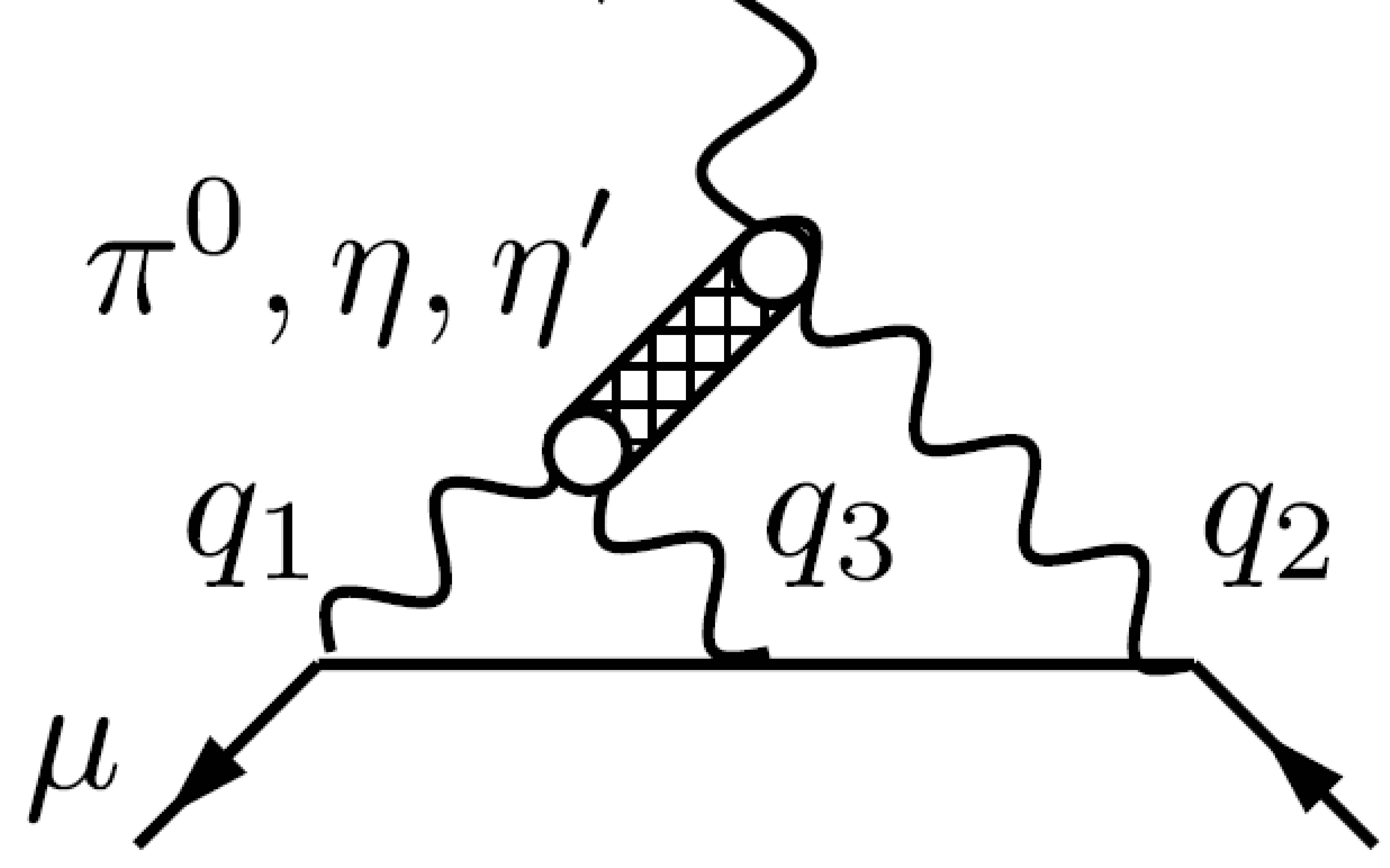}
 \caption{Left: 1st order contribution to $a$.
Center: Lowest-order hadronic VP diagram.
Right: A light-by-light diagram \cite{Jegerlehner:2009ry}.
 \label{fig:1}} 
 \end{center}
\end{figure}

Elementary particles have a magnetic moment $\vec{\mu}$ proportional
to their spin $\vec{s}$, with $\vec{\mu} = (g e)/(2m) \vec{s}$. 
While pointlike Dirac particles would have $g=2$, {\em i.e.} an
``anomalous'' relative deviation of $a\equiv (g-2)/2 = 0$, Nafe {\em et
al.} observed the first hints of a significant deviation from $a_e =
0$ more than 60 years ago \cite{Nafe:1947zz}.
The following year, Schwinger computed \cite{Schwinger:1948iu} the
first-order contribution to $a$, equal to $\alpha/(2\pi)$, the
diagram for which is shown in Fig.~\ref{fig:1}-left.

The development of  quantum electro-dynamics (QED) followed, and
later of gauge theories in general, making these early works the very
basis of our present understanding of the elementary world.
Tremendous efforts have been devoted to improving the precision of the
theoretical prediction and of the direct measurement of $a$ since
then \cite{Jegerlehner:2009ry}.

More than 60 years later the situation is pretty exciting, with the
experimental and theoretical precision on the anomalous magnetic
moment of the muon $a_\mu$ both of the order of $6. \times 10^{-10}$,
and a discrepancy of $(29 \pm 9) \times 10^{-10}$ between them,
{\em i.e.} amounting to 3.2 $\sigma$, should Gaussian statistics be assumed
(Table~\ref{tab:theory:exp}).

\begin{table}[h]
\small
\caption{Summary of the contribution to the theory prediction of the value of $a_\mu$, compared with the experimental measurement \cite{Jegerlehner:2009ry}.}
\begin{center} 
\begin{tabular}{lrrrr}
\hline
QED & 11 658 471.81 & $\pm$ 0.02 \\
Leading hadronic VP & 690.30 & $\pm$ 5.26 \\
Sub-leading hadronic VP & -10.03 & $\pm$ 0.11 \\
Hadronic light-by-light & 11.60 & $\pm$ 3.90 \\
Weak (incl. 2-loops) & 15.32 & $\pm$ 0.18 \\
\hline
Theory & 11 659 179.00 & $\pm$ 6.46 \\
Experiment \cite{Bennett:2006fi} & 11 659 208.00 & $\pm$ 6.30 \\
\hline
Exp $-$ theory & 29.00 & $\pm$ 9.03 \\
\hline
\end{tabular}
\end{center}
 \label{tab:theory:exp}
\end{table}

The largest contribution to $a_{\mu}$ by far is from QED but its
contribution to the uncertainty is negligible.
In terms of uncertainty, 
the main contribution is from the hadronic component of the one-loop
vacuum polarisation (VP, Fig.~\ref{fig:1}-center) 
and, to a lesser extent, from the hadronic component of the
light-by-light processes (Fig.~\ref{fig:1}-right).

The photon propagator with VP is obtained from the bare propagator by
 replacing the electric charge $e$ by the energy dependant quantity:
\begin{eqnarray*}
e^2 \rightarrow e^2 / [1 + (\Pi'(k^2) - \Pi'(0))], 
\end{eqnarray*}
where $k$ is the photon 4-momentum.
At low energy, hadronic processes are not computable with the desired
precision. Instead the VP amplitude $\Pi'(k^2)$ is obtained from the
dispersion relation:
\begin{eqnarray*}
\Pi'(k^2) - \Pi'(0) = \displaystyle\frac{k^2}{\pi} \int_{0}^{\infty} \displaystyle\frac{Im \Pi'(s)}{s(s - k^2 - i\epsilon)} \mbox{d} s, 
\end{eqnarray*}
which in turn is related through the optical theorem:
\begin{eqnarray*}
Im \Pi'(s) = \alpha(s) R_{\mbox{had}}(s) /3, 
\end{eqnarray*}
to the ratio:
\begin{eqnarray*}
R_{\mbox{had}}(s) = \sigma_{\mbox{had}} \displaystyle\frac{3s}{4\pi \alpha(s)} = 
 \displaystyle\frac{\sigma_{e^{+}e^{-}\to hadrons}}{\sigma_{e^{+}e^{-}\to \mu^{+}\mu^{-}}}. 
\end{eqnarray*}

\begin{figure}[t]
 \begin{center}
 \includegraphics[width=0.57\linewidth]{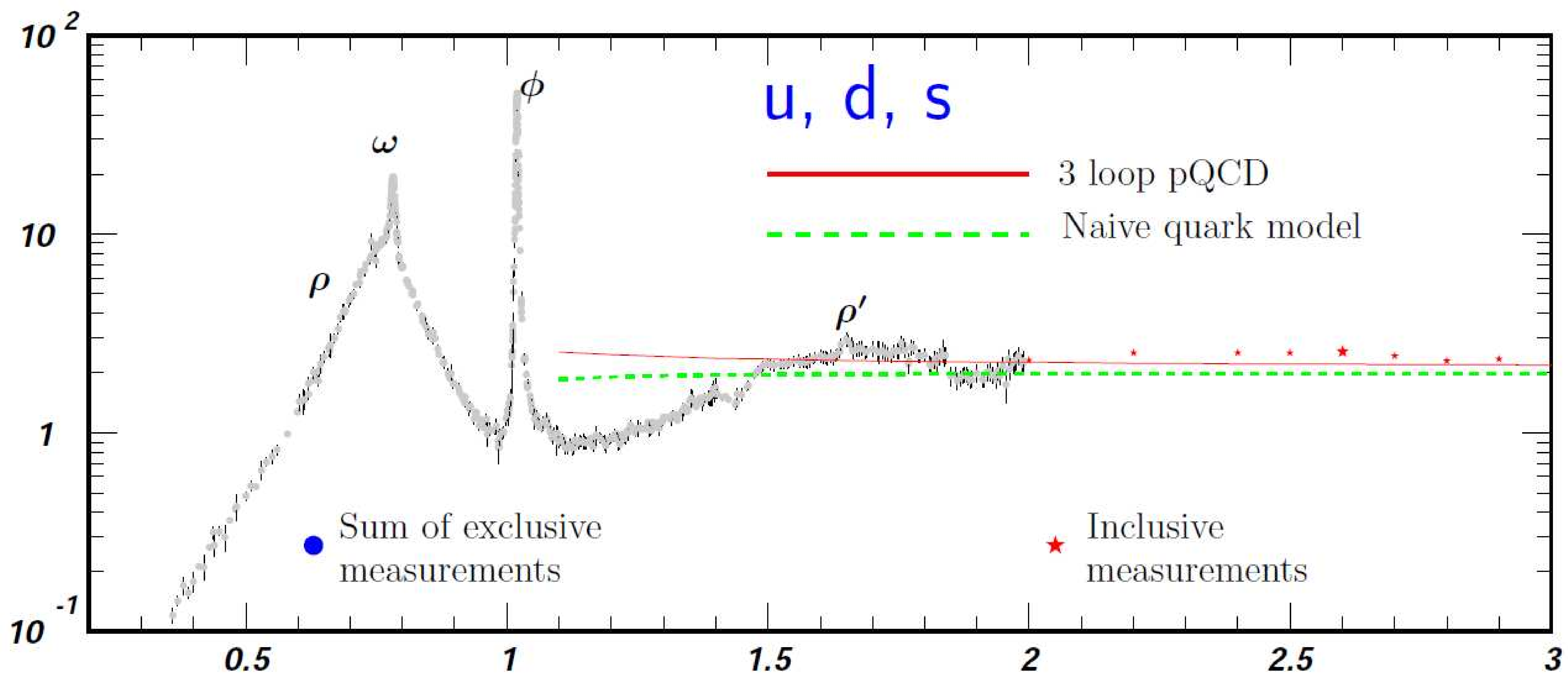}
 \includegraphics[width=0.40\linewidth]{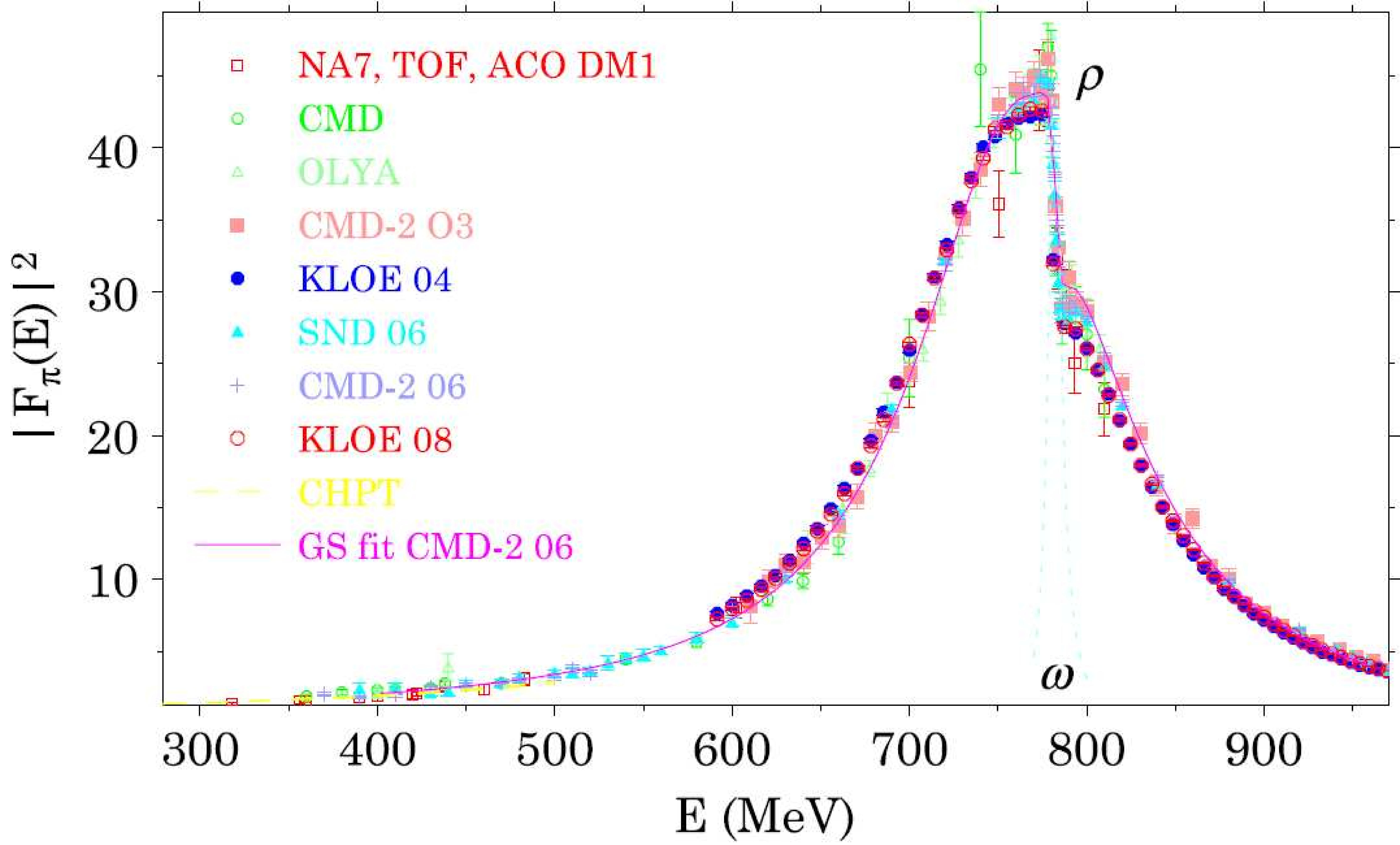}
 \caption{Left: 
$R_{\mbox{had}}$ as a function of $\sqrt{s}$ (GeV) \cite{Amsler:2008zzb}.
Right: 
Variation of the pion form factor squared with energy \cite{Jegerlehner:2009ry} (KLOE 04 is superseded by KLOE 08).
 \label{fig:pdg} }
 \end{center}
\end{figure}

Finally, the hadronic VP contribution is obtained from the
``dispersion integral'':
\begin{eqnarray*}
a_\mu^{\mbox{had}} = \left( \displaystyle\frac{ \alpha m_\mu} {3\pi} \right)^2
\int{ \displaystyle\frac{ R_{\mbox{had}}(s) \hat{K}(s)}{s^2} \mbox{d} s}, 
\end{eqnarray*}
where $\hat{K}(s)$ is an analytical function that takes values close to 1.
We note, from the $1/s^2$ variation of the integrand that the dominant
contribution comes from the low energy part of the integral.
A good experimental precision of the measurement of $R_{\mbox{had}}(s)$
at low energy is therefore welcome.
Fig.~\ref{fig:pdg}-right shows a summary of the present measurements of
$R_{\mbox{had}}(s)$ \cite{Amsler:2008zzb}, where the presence of 
$J^{PC} = 1^{--}$ mesons can be seen.

The $\pi^{+}\pi^{-}$ channel has both the largest contribution and 
dominates the uncertainty, with 
$a_{\mu}^{\pi^{+}\pi^{-}}[2m_\pi, 1.8 \mbox{GeV}/c^2] = 
(504.6 \pm 3.1 (\mbox{exp}) \pm 0.9 (\mbox{rad}))\times 10^{-10}$,
compared to the full 
$a_\mu^{\mbox{had}} = (690.9 \pm 5.3)\times 10^{-10}$ from Table
\ref{tab:theory:exp}.
A summary of direct measurements in terms of the squared pion form
factor is shown in Fig.~\ref{fig:pdg}-left.
The 3.2 $\sigma$ discrepancy mentioned above is computed using this
$\pi\pi$ input.

\section{$\tau$ decay spectral functions}

The $I=1$ part of the $e^{+}e^{-} \to \pi^{+}\pi^{-}$ cross-section
can be estimated from the spectral function of $\tau$ decays to $\nu
\pi^{+}\pi^{0}$, under the hypothesis of conservation of the weak
vector current (CVC).
The method was pioneered by the ALEPH collaboration 
\cite{Barate:1997hv} and followed by OPAL 
\cite{Ackerstaff:1998yj}
and CLEO 
\cite{Anderson:1999ui}. 
Recently the Belle collaboration has performed an analysis using 
much larger statistics \cite{Fujikawa:2008ma}, obtaining a value
compatible with, and more precise than, the combination of all
previous results
\cite{Davier:2002dy}.

The $\tau$ method provides a high experimental precision, but 
extracting the contribution to $a_\mu$ depends on making a number
of isospin-breaking (IB) corrections. 
A recent update \cite{Davier:2009ag} of 
 \cite{Davier:2002dy} lowers the correction by $\approx
7 \times 10^{-10}$, while the uncertainty on the correction is now
$1.5 \times 10^{-10}$. 

The branching fraction of the $\tau \to \nu \pi^{+}\pi^{0}$
decay also takes part in the calculation, with a 0.5 \% uncertainty.

\section{$e^{+}e^{-} \to \pi^{+}\pi^{-}$ using ISR method}

Initial-state radiation (ISR) makes it possible to measure the cross-section
of the production of a final state $f$ in $e^{+}e^{-}$ collisions at a
squared energy $s$, over a wide range of energies, lower than
$\sqrt{s}$, through the radiation of a high energy photon by one of the
incoming electrons, after which the electrons collide at a squared
energy $s'$.

The BaBar experiment has developed a systematic program of
measurements of cross-sections of $e^{+}e^{-}$ to hadrons at low energy
using the ISR method \cite{babar:isr}.
The boost undergone by the final state $f$ provides an excellent
efficiency down to threshold.
In all studies by BaBar, the ISR photon is observed ($\gamma$-tag) and
its direction is compared to the direction predicted from the
direction of $f$, providing a powerful rejection of background noise.
Most of these measurements are more precise than the
previously available results by about a factor of three.

\subsection{KLOE's result on $e^{+}e^{-} \to \pi^{+}\pi^{-}$} 

The KLOE experiment, when running on the $\phi$ resonance,  studied the
$e^{+}e^{-}$ annihilations to $ \pi^{+}\pi^{-}$ with the ISR method
\cite{:2008en}.
Here the ISR photon is not reconstructed: the requirement that the photon
direction be compatible with its having been emitted in the beam pipe
allows mitigation of the background to some extent, but the
systematical uncertainty on background subtraction is still a major
component of the total uncertainty.
The radiator function is provided from simulation, with systematics of
0.5\%, and is the other major component.

The value of $a_{\mu}^{\pi^{+}\pi^{-}}$ obtained is compatible with
the combination of previous results by CMD-2 \& SND over the mass
 range that they have in common of (630 - 958 MeV/$c^2$).

\subsection{BaBar's result on $e^{+}e^{-} \to \pi^{+}\pi^{-}$} 

BaBar uses a different approach: photon tagging 
with the ISR luminosity obtained from the muon channel, 
 $e^{+}e^{-} \to \mu^{+}\mu^{-} \gamma$ \cite{Aubert:2009fg}.
The systematics related to additional radiation is minimized in this
NLO measurement, {\em i.e.} radiation of one possible additional photon is
allowed, so that the final states actually reconstructed are
 $\pi^{+}\pi^{-} \gamma(\gamma)$ and 
 $\mu^{+}\mu^{-} \gamma(\gamma)$.
The ``bare'' ratio $R_{\mbox{had}}(s')$ mentioned above is obtained from the
experimentally measured $R_{exp}(s')$ after correction of final state
radiation (FSR) in
$e^{+}e^{-} \to \mu^{+}\mu^{-}$ and of additional FSR in ISR events 
 $e^{+}e^{-} \to \mu^{+}\mu^{-} \gamma $.

A number of important systematics cancel when measuring the $\pi/\mu$
ratio, such as those associated with the collider luminosity, the
efficiency of the reconstruction of the ISR photon, and the
understanding of additional ISR radiation.

The limiting factor is then the understanding of the possible
``double'' $\pi-\mu$, MC-data efficiency discrepancies. These are
studied in detail, with methods designed to disentangle correlations
as much as possible.
For example, inefficiency of the track-based trigger is studied using
events selected with a calorimetry-based trigger -- the small
correlation between both triggers being studied separately.
Likewise, $\mu$ and $\pi$ particle identification (PID) efficiency is
studied in good-quality, two-track ISR events, in which either one,
or both, tracks meet the PID selection criteria.
Concerning tracking, a sizable degradation of the efficiency for
tracks overlapping in the detector was observed and studied in detail.

The systematics finally obtained are of the order of, or smaller than,
1 \% over the whole mass range studied; {\em i.e.}, from threshold to 3
GeV/$c^2$.

\begin{figure}[t]
 \begin{center}
 \includegraphics[width=0.95\linewidth]{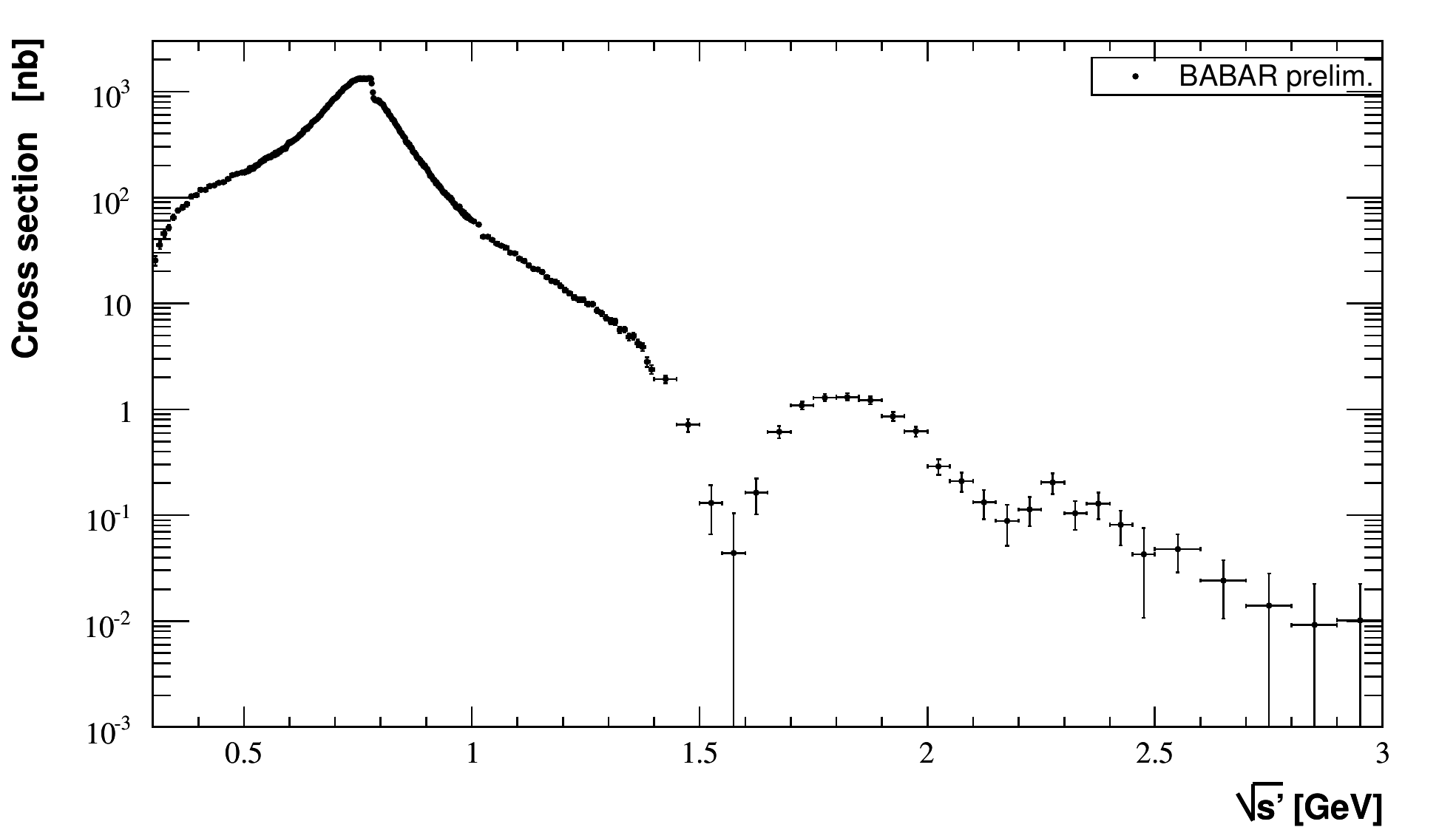}
 \caption{Bare, unfolded 
\cite{Malaescu:2009dm}, 
$e^{+}e^{-} \to \pi^{+}\pi^{-}$ cross section measured by BaBar, using the ISR method \cite{Aubert:2009fg}.}
 \label{fig:babar:spectre} 
 \end{center}
\end{figure}

The $e^{+}e^{-} \to \pi^{+}\pi^{-}$ cross section measured by BaBar is
shown in Fig.~\ref{fig:babar:spectre}. The sharp drop due to the
interference between the $\rho$ and the $\omega$ is clearly visible.
The interference between the successive radial excitations of the
$\rho$ induces these dips in the cross-section.

The measured value of $a_{\mu}^{\pi^{+}\pi^{-}}[2m_\pi, 1.8\mbox{GeV}/c^2] =$
$ (514.1 \pm 2.2 \pm 3.1) \times 10^{-10}$
has a precision similar to the combination of all previous
$e^{+}e^{-}$-based results, but is larger by about 2.0 $\sigma$.

In addition to the measurement of the $\pi/\mu$ ratio, and extraction
of the $e^{+}e^{-} \to \pi^{+}\pi^{-}$ cross section, BaBar has
compared its $\mu^{+}\mu^{-}$ spectrum to the Monte Carlo prediction,
finding a good agreement within $0.4 \pm 1.1 \%$, dominated by the
collider luminosity uncertainty of $\pm 0.9 \%$.

\begin{figure}[t]
\begin{center}\small
\begin{tabular}{ccccccc} 
Belle $\tau$ & CMD2 $e^{+}e^{-}$ \\
\includegraphics[width=0.49\linewidth]{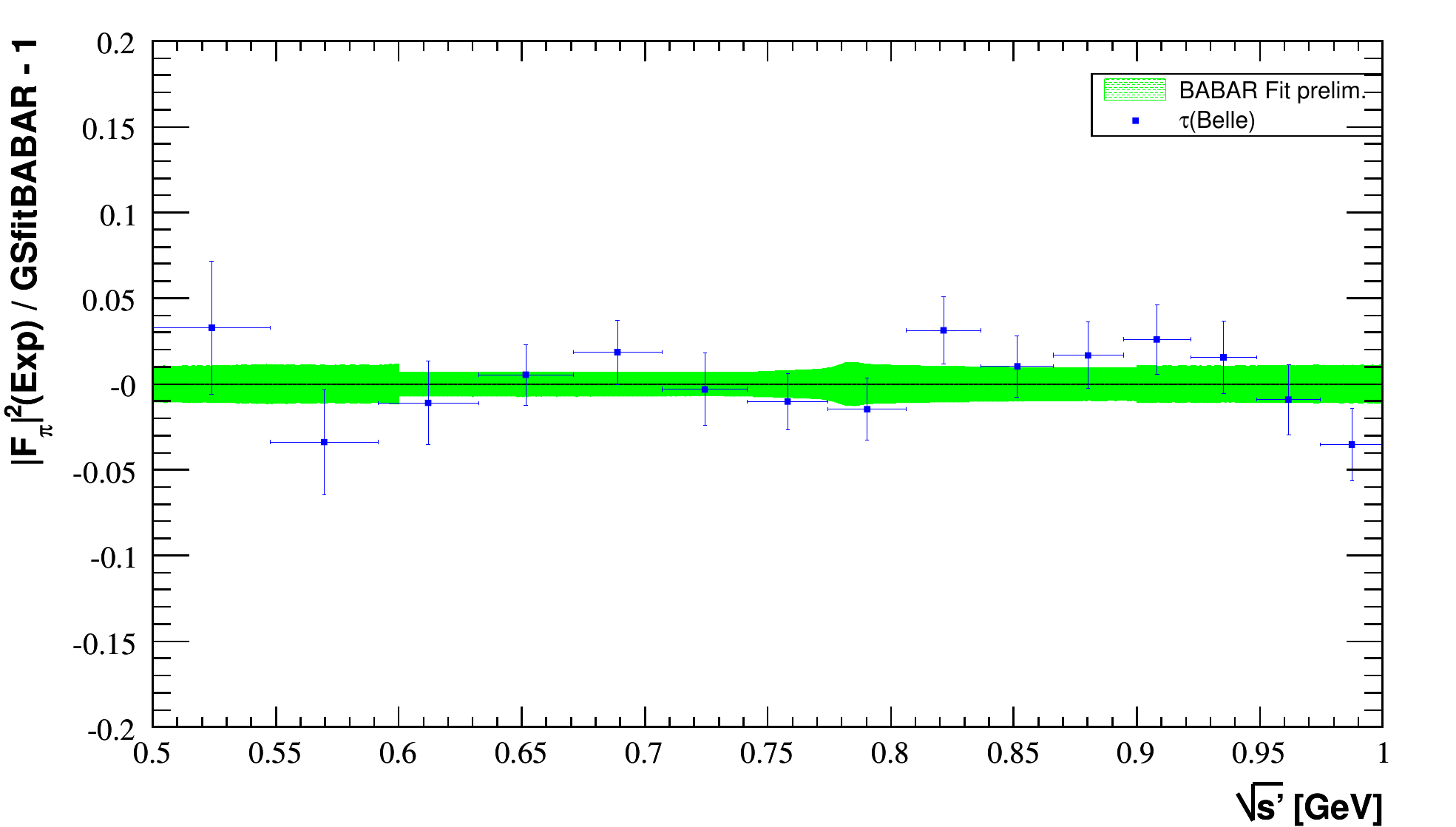}
&
\includegraphics[width=0.49\linewidth]{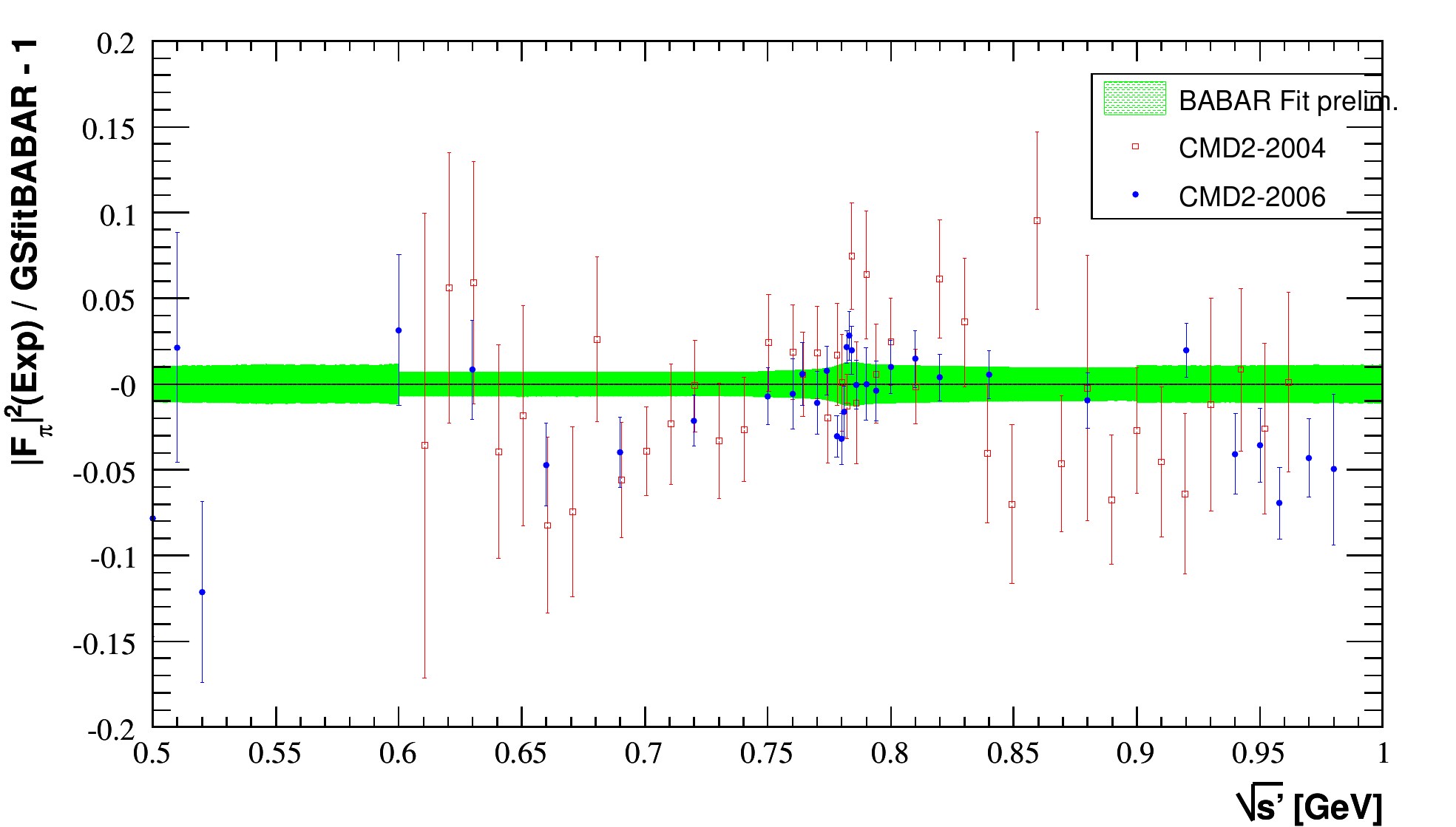}
\\
KLOE $e^{+}e^{-}$ ISR & SND $e^{+}e^{-}$ \\
\includegraphics[width=0.49\linewidth]{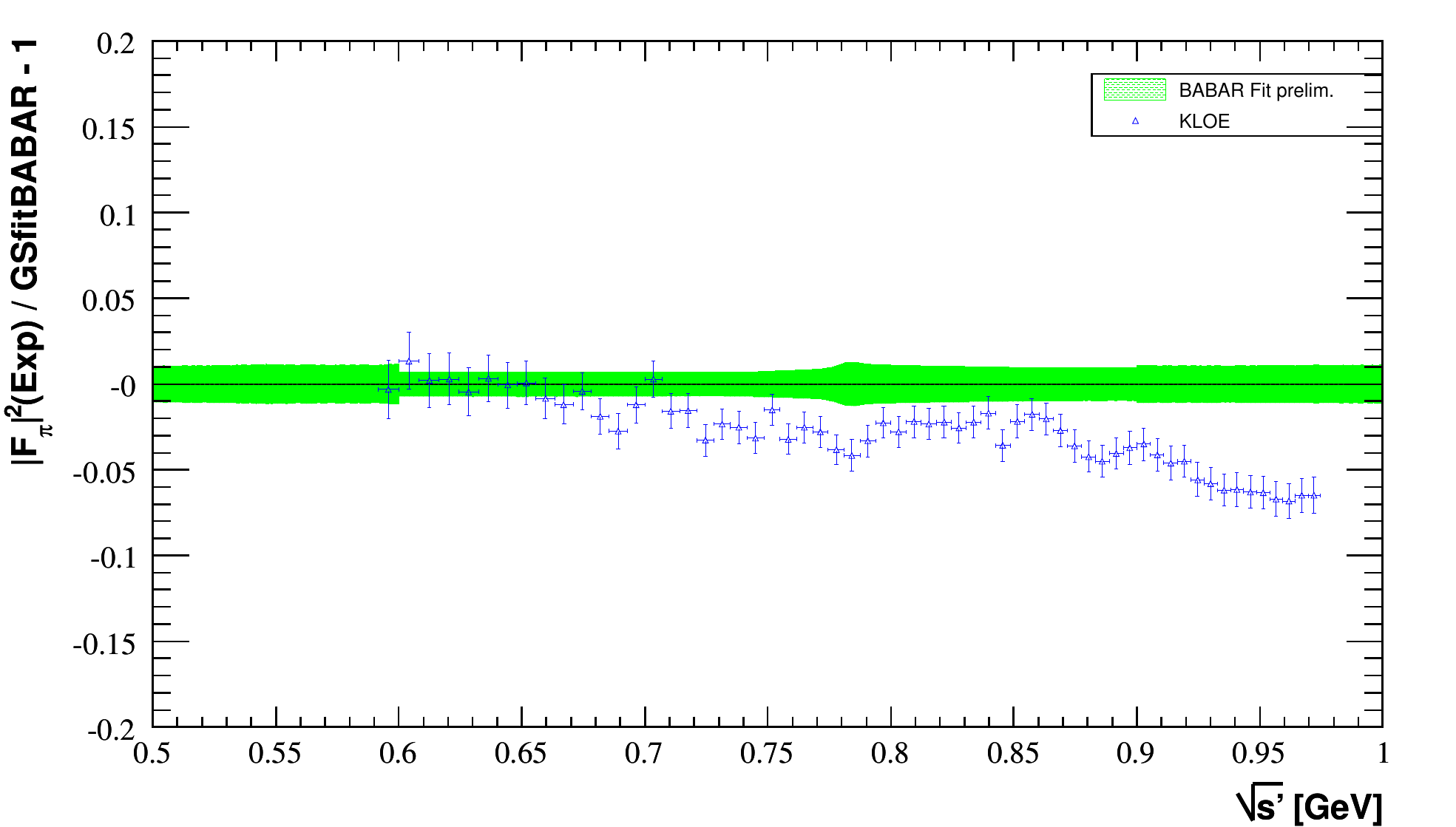}
&
\includegraphics[width=0.49\linewidth]{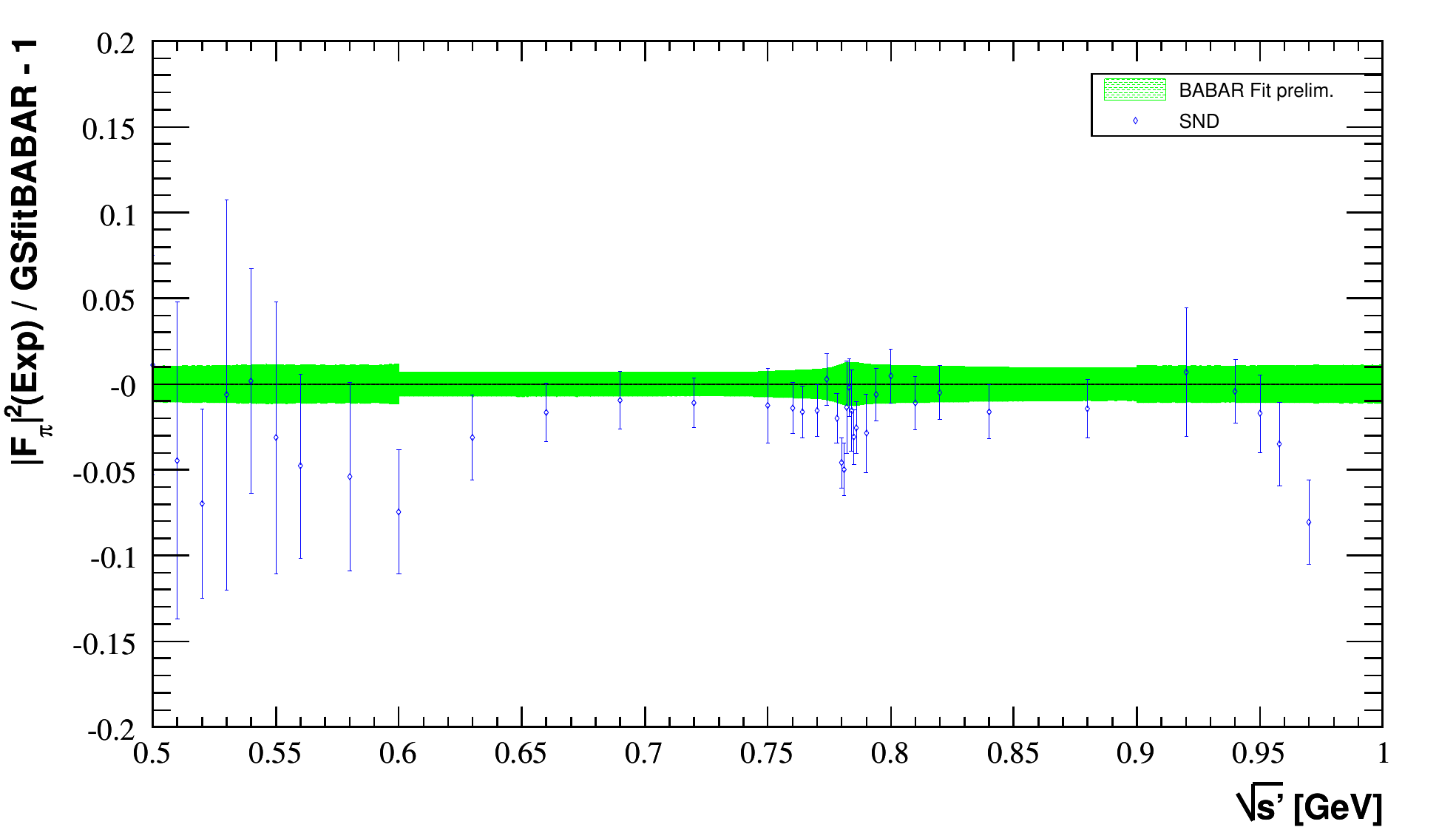}
\end{tabular}
 \caption{Relative difference between the BaBar result with that of previous experiments.}
 \label{fig:babar:comp} 
 \end{center}
\end{figure}

The distribution of the squared pion form-factor is fitted with a
vector-dominance model including the resonances $\rho, \rho', \rho'',
\omega$, with the $\rho$'s being described by the Gounaris-Sakurai
model.
The fit (Figure in Ref. \cite{Bernard:2009nb}) yields a good $\chi^2 /n_{df}$ 
of $334 / 323$, and parameters compatible with the world-average
values.
BaBar can then use the fitted model to compare their result with that
 of previous measurements (Fig.~\ref{fig:babar:comp}).
The BaBar result is a bit larger than that obtained by CMD2 \cite{Aulchenko:2006na} and SND 
\cite{Achasov:2005rg},
nicely compatible with the high-statistics $\tau$-based result by
Belle, but shows a clear disagreament with KLOE.

\section{$a_{\mu}$: the present situation}

The present situation in terms of $a_{\mu}$ is summarized in 
Fig.~\ref{fig:babar:situation}:
\begin{figure}[t]
 \begin{center}
 \includegraphics[width=0.95\linewidth]{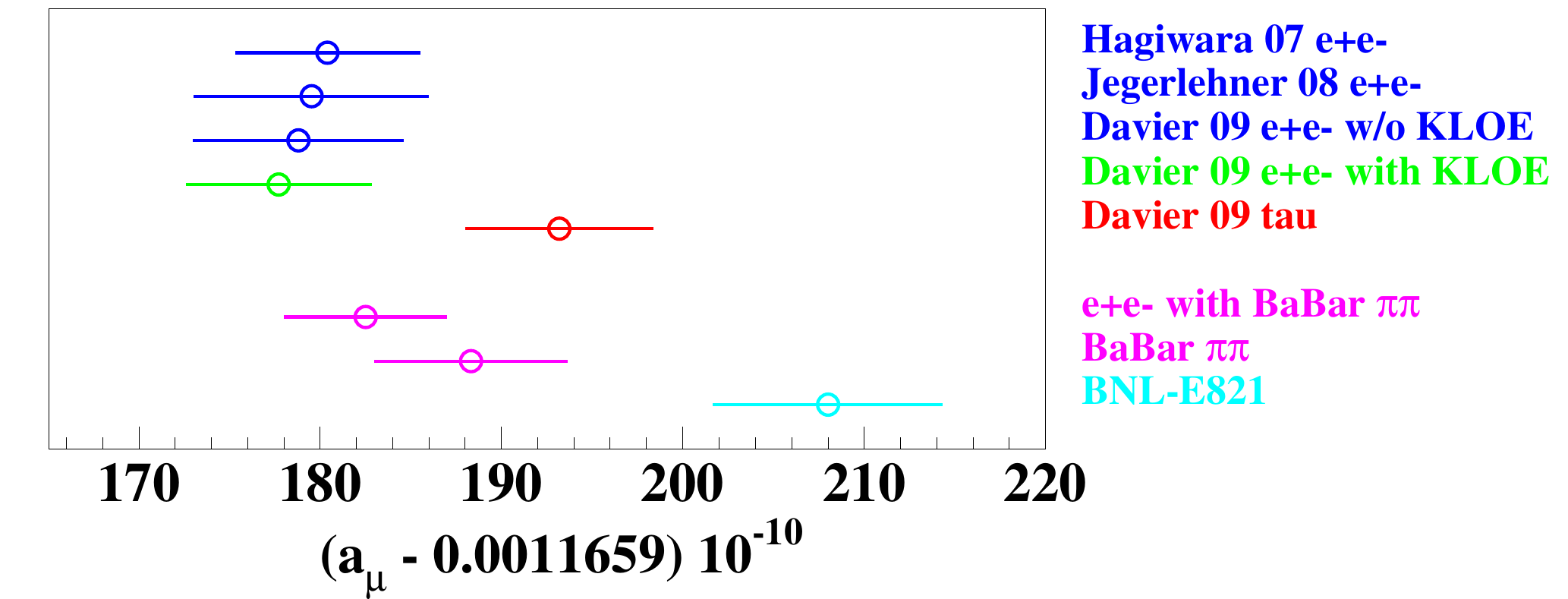}
 \caption{$a_{\mu}$: the present situation.}
 \label{fig:babar:situation} 
 \end{center}
\end{figure}

\begin{itemize}
\item
The four upper points show that there is a general agreement between
 the various recent combinations of direct $e^{+}e^{-}$-based
 $\pi^{+}\pi^{-}$ measurement
\cite{Hagiwara:2006jt,Jegerlehner:2008zz,Davier:2009ag}.
\item
The large discrepancy between computations of $a_\mu$ based on these and the experimental measurement
by BNL-E821 \cite{Bennett:2006fi} is clear.
\item
The combination of $\tau$-based results, when corrected for
isospin-breaking effects using the most recent
calculation \cite{Davier:2009ag}, are also significantly lower than the
experimental measurement \cite{Bennett:2006fi}, by 1.8 $\sigma$.
\item
My computation of $a_{\mu}$ using the BaBar $\pi^{+}\pi^{-}$
measurement \cite{Aubert:2009fg} only is larger than the combination
of previous $e^{+}e^{-}$-based measurements, and compatible with the
$\tau$-based result, but still 2.4 $\sigma$ away from BNL-E821
\cite{Bennett:2006fi}.
\item 
The combination of all $e^{+}e^{-}$-based measurements, including the
recent one by BaBar, shows an uncertainty that has decreased
significantly, and a central value that is larger.
However, the significance of the difference with respect to BNL-E821
 \cite{Bennett:2006fi} is barely changed, of the order of 3.3 $\sigma$.
\end{itemize}

A more sophiticated combination of the available results, published
 recently \cite{Davier:2009zi}, yields similar numbers.

\section{What might take place during this decade}

\subsection{$a_\mu$ measurement}

One single high-precision statistics-dominated experimental
measurement of $a_{\mu}$ \cite{Bennett:2006fi} is facing a prediction
in which the contribution with largest uncertainty has been 
confirmed, within reasonable significance, by a number of measurements
using various methods, each affected by its own systematics.

The obvious next step, before calling for new physics, is therefore to
check the measurement:
\begin{itemize}
\item A new
collaboration is planning to move the experimental apparatus from BNL
to FNAL, and perform a new measurement with statistics increased by
a factor of 50, and reduced systematics, bringing the experimental
uncertainty down to 0.14 ppm, {\em i.e.}, $1.6 \times 10^{-10}$
 \cite{Carey:2009zz}.
\item 
It would obviously be intensely desirable to cross-check such
a measurement using a completely different set-up. An alternative
 scheme is explored at J-PARC, with a micro-emittance muon beam
 inside a high-precision magnetic field, mono-magnet storage ``ring''
 \cite{Naohito:Saito}.
\end{itemize}

\subsection{Prediction}

On the prediction side, the main effort is understandably devoted to
the hadronic VP contribution.
\begin{itemize}
\item BaBar will complete its ISR program and
provide measurements of all possible hadronic final states in the low
energy range relevant to this discussion.
\item 
Belle may check BaBar's $\pi^{+}\pi^{-}$ measurement and BaBar may
check Belle's $\tau$ spectral functions. KLOE is working on an
analysis with photon tagging too.
\item 
BES-III will measure $R_{\mbox{had}}(s)$ in the range 2.0 -- 4.6 GeV,
something that will improve on $a_{\mu}$ only marginally, but will
also measure the $\tau \to \nu \pi^{+}\pi^{0}$ branching fraction with
improved precision \cite{Zhemchugov:2009zz}, an important ingredient
in the use of the $\tau$-based spectral functions.
\item 
The recent calculation of isospin-breaking corrections
\cite{Davier:2009ag} will doubtlessly be
cross-checked by other authors.
\item 
The collider at Novosibirsk has been upgraded to VEPP-2000
\cite{Shatunov:2008zz}, and the CMD 
\cite{Fedotovich:2006cq}
and SND experiments too.
\end{itemize}

Following the vacuum polarisation, the next target in line for
improvement is the contribution of light-by-light scattering. Here
too work is in progress and there is hope to improve the precision,
both theoretically \cite{Prades:2008zz}, and using results of the
$\gamma\gamma$ programe at $DA\Phi NE-2$ \cite{Archilli:2008zz}.

~

In total, there is good hope to bring both the prediction and
experimental uncertainties of $a_{\mu}$ at a very few $10^{-10}$.

I regret I didn't have the time to present the implications of the
$a_{\mu}$ discrepancy, if assumed to be due to an underestimated
hadronic cross-section, on the estimation of the Higgs mass 
\cite{Passera:2008jk}.
Interpretation of the discrepancy as being due to contribution of
yet-unknown heavy object(s) in loops is also an interesting
possibility \cite{Stockinger:2006zn}.

Finally, at higher energy, ISR can be used to understand QCD by 
exploring the new spectroscopy of $J^{PC} = 1^{--}$ charmonium-like
states, opened by the discovery by BaBar of the $Y(4260)$ meson
\cite{Aubert:2005rm}.

\end{document}